\documentclass[12pt]{iopart}
\usepackage[separate-uncertainty = true,per-mode=reciprocal,range-units=single]{siunitx}
\usepackage{cite}
\expandafter\let\csname equation*\endcsname\relax
\expandafter\let\csname endequation*\endcsname\relax
\usepackage[version=3]{mhchem}
\usepackage{graphicx}
\usepackage{subfigure}
\usepackage{todonotes}
\usepackage{amsmath,mathtools}
\usepackage{url}
\usepackage{booktabs}

\begin{document}
\title[Exploring the electron density in EUV-induced argon plasmas]{Exploring the electron density in plasmas induced by extreme ultraviolet radiation in argon}
\author{R M van der Horst$^1$, J Beckers$^1$, E A Osorio$^2$ and V Y Banine$^{1,2}$}
\ead{r.m.v.d.horst@tue.nl}
\address{$^1$ Department of Applied Physics, Eindhoven University of Technology, PO Box 513, 5600MB Eindhoven, The Netherlands}
\address{$^2$ ASML The Netherlands B.V., PO Box 324, 5500AH Veldhoven, The Netherlands}

\begin{abstract}
The new generation of lithography tools use high energy EUV radiation which ionizes the present background gas due to photoionization. To predict and understand the long term impact on the highly delicate mirrors It is essential to characterize these kinds of EUV-induced plasmas. We measured the electron density evolution in argon gas during and just after irradiation by a short pulse of EUV light at 13.5 nm by applying microwave cavity resonance spectroscopy. Dependencies on EUV pulse energy and gas pressure have been explored over a range relevant for industrial applications. 

Our experimental results show that the maximum reached electron density depends linearly on pulse energy. A quadratic dependence – caused by photoionization and subsequent electron impact ionization by free electrons - is found from experiments where the gas pressure is varied. This is demonstrated by our theoretical estimates presented in this manuscript as well.
\end{abstract}

\submitto{\JPD}

\maketitle

\section{Introduction}

There is an increasing global demand for more computational power and memory capacity at lower cost and higher energy efficiency. The multibillion-dollar semiconductor industry is struggling to meet this demand and this is why they continuously aim to further miniaturize computer chips. To achieve this further miniaturization, the upcoming generation of commercial lithography tools use extreme ultraviolet (EUV) radiation at a wavelength of \SI{13.5}{\nano\meter}, being equivalent to a photon energy of \SI{92}{\electronvolt}. This EUV radiation will be generated by a pulsed light source based on plasmas in tin vapour, which are generated by irradiating liquid tin droplets with an intense laser beam~\cite{Bergmann2001,Jonkers2006}. The absorption cross section of EUV radiation in gasses is very high~\cite{Saito1992}, so ideally the lithography tools would operate at high vacuum. However, for technical reasons manufacturers are forced to operate the tool with a low background gas (\SIrange{0.1}{30}{\pascal}). Consequently, the interaction between the EUV photons and the background gas generates a plasma by photo-ionization, i.e. an EUV-induced plasma, everywhere it travels. In the vicinity of surfaces, the plasma induces strong electric fields due to the formation of a plasma sheath. Typically, the thickness of these sheaths is in the order of \SIrange{5}{10}{\milli\meter} and the electric fields are in the order of \SI{~e4}{\volt\per\meter}~\cite{Beckers2011}. These strong electric fields accelerate ions to high velocities towards the highly delicate and expensive EUV multilayer optics, which can have a long term impact on their operation~\cite{VanderVelden2006}. Eventually, controlling this long term impact requires more understanding of the fundamental processes in EUV-induced plasmas.

Previous research used numerical simulations to investigate the fundamental properties of the plasma and its effect on optical elements~\cite{VanderVelden2006,VanderVelden2008}. In that study the authors also attempted to measure the electron density --being one of the key plasma parameters-- using Langmuir probes; however, they concluded that the used probes are not feasible~\cite{VanderVelden2008}. In ref.~\cite{VanderHorst2014} we reported for the first time non-intrusive measurements of the electron density in an EUV-induced plasma in argon. For these measurements we applied a method known as microwave cavity resonance spectroscopy~\cite{Beckers2009,VandeWetering2012,VanderHorst2014,Stoffels1995,Gundermann2001}. Where our previous work focused on solely the time evolution of the electron density and characterization of the MCRS method, the current paper explores the density over a range of parameters representable to those used in commercial lithography tools.

The goal of this research is to answer the following question:
How does the density of electrons in a plasma generated by pulsed EUV-radiation depend on gas pressure and the intensity of the EUV radiation?

\section{Microwave cavity resonance spectroscopy}
The applied diagnostics to measure the electron density is a method known as microwave cavity resonance spectroscopy (MCRS). This method is extensively used to measure the electron density in other types of plasmas~\cite{Beckers2009,VandeWetering2012,Stoffels1995,Gundermann2001}, while, recently, we published results of the first application of MCRS to EUV-induced plasmas~\cite{VanderHorst2014}. Since the MCRS technique is described in great detail in these publications, we will suffice with only briefly describing the key concepts here. In MCRS measurements a standing wave is excited in a (cylindrical) resonance cavity. This wave only exists at specific frequencies, called the resonant frequencies $\omega_0$, which depend, amongst others, on the permittivity of the medium inside the cavity. When free electrons are created in the cavity, the permittivity changes, due to which the resonant frequency will shift with $\Delta\omega$. From this $\Delta\omega$ the electron density can be determined~\cite{VandeWetering2012}:
	\begin{equation}
	\bar{n}_e = \frac{2m_e \varepsilon_0}{e^2}
	    \frac{\omega^2}{\omega_0} \Delta\omega
	\end{equation}
with $m_e$ the electron mass, $e$ the elementary charge and $\omega_0$ and $\omega$ the resonant frequencies without and with plasma, respectively. Note that this method gives the averaged electron density weighted with the square of the local electric field of the standing wave of the excited mode $\vec{E}(\vec{r})$~\cite{VandeWetering2012}:
	\begin{equation}
	\bar{n}_e = \frac{\int_{cavity} n_e(\vec{r}) E^2(\vec{r}) d\vec{r}}
		{\int_{cavity} E^2(\vec{r}) d\vec{r}}.
	\label{eq:avDensity}
	\end{equation}
To determine the local electron density, the spatial profile on the electron density needs to be known. Since this profile is unknown, we will present only the square-electric-field weighted averaged electron density.

\section{Experimental set-up}
We used a xenon-based EUV source to generate EUV radiation~\cite{Bergmann2001}. The experimental details of this source and the diagnostics will be discussed in this section.

\subsection{EUV source}
\begin{figure}
	\hfill
	\subfigure[]{
		\raisebox{-0.5\height}{\includegraphics{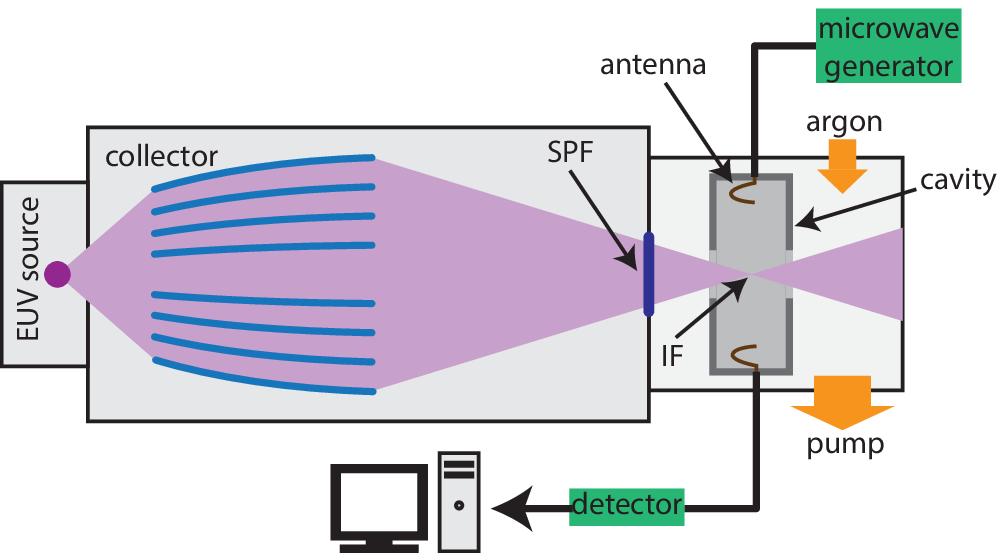}}
	}
	\hfill
	\subfigure[]{
		\raisebox{-0.5\height}{\includegraphics{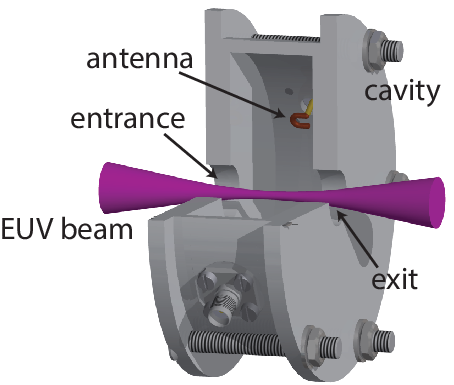}}
	}
	\hfill
	\caption{(a) Schematic drawing of the experimental set-up. The EUV radiation is created in the source chamber by a xenon pinch discharge. The radiation is collected by a set of elliptic mirrors and focused in the measurement chamber. The cavity is placed in the focus. The gas environment of the measurement chamber is screened from the collector chamber by a spectral purify filter (SPF). This filter transmits between approximately 10-20 nm. The argon pressure in the measurement chamber is varied with a needle valve. (b) Drawing of the microwave cavity with an indication of the EUV beam~\cite{VanderHorst2014}.}
	\label{fig:setup}
\end{figure}

The pulsed EUV radiation, with a pulse duration of \SI{150}{\nano\second} and a repetition rate of \SI{500}{\hertz}, is generated by a pulsed xenon pinch plasma in the source chamber (see Figure~\ref{fig:setup})~\cite{Bergmann1999}. The radiation is collected by a set of elliptic multilayer mirrors in the collector chamber, which focussed the EUV radiation in the intermediate focus (IF) in the measurement chamber. The EUV beam had a waist of \SI{4}{\milli\meter} and a divergence of \SI{10}{\degree}. To allow high vacuum in the measurement chamber, the collector and measurement chamber are differentially pumped and separated by an aperture of \SI{4}{\milli\meter}. In front of the aperture, we placed a spectral purity filter (SPF), which transmits between only \SI{10}{\nano\meter} and \SI{20}{\nano\meter} and, as such, prevents out-of-band radiation in the measurement chamber.

\subsection{Measurement vessel}
We placed diagnostics to study the induced plasma in the measurement vessel. As mentioned earlier, the main diagnostic technique was microwave cavity resonance spectroscopy. In addition to that, we determined the temporal shape of the EUV pulse from the secondary electron emission from a copper disk. These techniques will be described in more detail below. 

\begin{figure}
		\centering
		\includegraphics{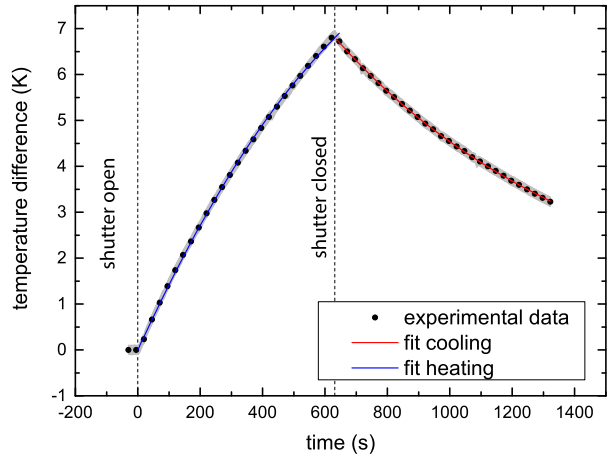}
		\caption{Heating and cooling of the temperature sensor (every 5th data point is shown to improve readability). The heat loss terms are determined from the cooling curve ($C_1 = \SI{9.1+-0.1e-4}{\watt\per\kelvin}$ and $C_2 = \SI{3+-2e-5}{\watt\per\square\kelvin}$). These terms are used to fit the heating curve with equation~\ref{eq:heat_loss}. The EUV power is \SI{0.026+-0.001}{\watt}.}
		\label{fig:temperature}
	\end{figure}

We determined the spectrally integrated EUV power with a temperature sensor which heats up due to the absorption of EUV radiation. The temporal temperature increase is a measure for the time-averaged EUV power. The sensor consists of a copper disk (diameter of \SI{25.4}{\milli\meter} and thickness of \SI{1.0}{\milli\meter}) and a temperature transducer with a repeatability of \SI{+-0.1}{\celsius}. The temperature of the sensor is affected by the EUV power $P_{in}$, radiative losses and conductive losses. $\tfrac{d}{dt}\Delta T$ is then given by:
	\begin{equation}
	\frac{d}{dt}\Delta T \approx \frac{1}{m_c c_p}\left(P_{in} - 
		(\underbrace{4\sigma A_d T_0^3 + \lambda_c}_{C_1})\Delta T - \underbrace{6\sigma A_d T_0^2}_{C_2} \Delta T^2\right),
	\label{eq:heat_loss}
	\end{equation}
where $m_c$ is the mass of the copper disk, $c_p$ is the specific heat capacity of copper, $\sigma$ is the Stefan-Boltzmann constant, $A_d$ is the area of the copper disk, $T_0$ is the temperature of the environment and $\lambda_c$ is the conductive heat loss coefficient. During a power measurement, we first expose the power sensor to EUV radiation, due to which the temperature increases. After a few minutes of exposure, we close a shutter and the power sensor cools down. The resulting temperature curve is shown in Figure~\ref{fig:temperature}. We determined the heat loss terms by fitting the temperature decrease with equation~\ref{eq:heat_loss} ($P_{in}=0$). With these -now known- loss terms we can fit the temperature increase and obtain the EUV power. The error in the power determination is \SI{5}{\percent}. In the case of the sensor response of Figure~\ref{fig:temperature}, the power is \SI{0.026+-0.001}{\watt}, which means an EUV pulse energy of \SI{53+-3}{\micro\joule} at a repetition rate of \SI{500}{\hertz}. 

To measure the temporal evolution of the radiation energy during an EUV pulse, we replaced the temperature sensor (copper disk and temperature transducer) by a copper disk only. This disk is connected to ground via a \SI{50}{\ohm}-resistor and we measured the voltage over this resistor with an oscilloscope. Due to the photo-electric effect, the EUV radiation induces a current. This is why, the temporal evolution of the current is representative for the temporal evolution of the EUV pulse.

We varied the argon pressure in the measurement vessel between \SIlist{0.1;20}{\pascal} with a needle valve. 

\subsection{MCRS set-up}
\begin{figure}
	\centering
	\includegraphics{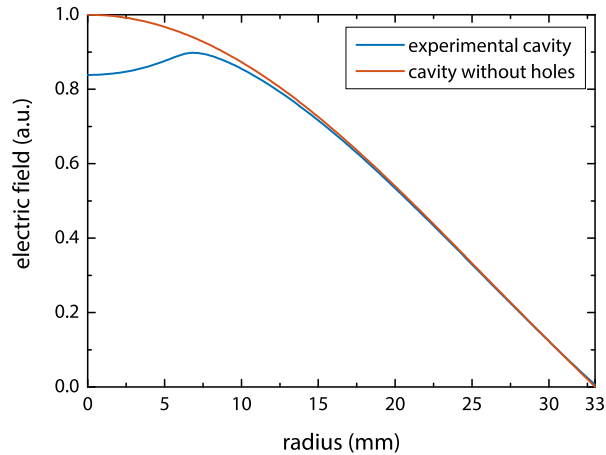}
	\caption{Modelled electric field of the TM$_{010}$ mode in the experimental cavity~\cite{VanderHorst2014}. The electric field is calculated with the plasimo platform~\cite{VanDijk2009,Diaz2011}. Moveover, the electric field of an ideal cavity without holes is shown.}
	\label{fig:Efield}
\end{figure}

We placed the microwave resonant cavity around the intermediate focus in the measurement vessel. This cavity is made of aluminium and has an inner radius of \SI{33}{\milli\meter} and an inner height of \SI{20}{\milli\meter}. It has entrance and exit holes for the EUV beam with radii of \SI{6.5}{\milli\meter}. Inside the cavity two copper antennas are placed in the side-walls at opposite sides, one functioning as transmitter, the other as receiver. These antennas do not face the EUV light directly. A low power (\SI{10}{\milli\watt}) microwave generator (Stanford Research Systems SG386) excited the TM$_{010}$ mode at \SI{3.49674+-0.00004}{\giga\hertz}. We calculated the electric field of the TM$_{010}$ mode in the exact cavity geometry with the plasimo platform~\cite{VanDijk2009,Diaz2011} and the result is shown, along with the same resonant mode in a similar cavity without holes, in Figure~\ref{fig:Efield}. This graph shows that the entrance and exit hole have a small influence on the electric field profile of the excited mode. The electric field is maximum in the centre of the cavity. This means that the electron density is most effectively sampled in the centre of the cavity (see equation~\ref{eq:avDensity}), which is also the position where the plasma is created during and just after the EUV pulse when the narrow EUV beam is directed axially through the centre of the cavity. The response of the cavity is measured with a microwave detector (Hittiti HMC602LP4) which has a time response of \SI{10}{\nano\second}. The resonant frequency is determined by sweeping the microwave generator over a frequency range and measuring the response. At every point in time we fit the cavity response as function of the microwave frequency with a (two component) Fourier series. The frequency at which this fit is maximum, is the resonant frequency. It has been shown in~\cite{VanderHorst2014} that the response time of the cavity is \SI{14}{\nano\second} and that the combined response time of the cavity and the detector is about \SI{17}{\nano\second}. The error in the square-electric-field weighted averaged electron density is less than \SI{30}{\percent} with a detection limit of \SI{2E12}{\per\cubic\meter}, as has been shown as well in~\cite{VanderHorst2014}.

\section{Results}
\subsection{Gas pressure}
\begin{figure}[!h]
  \centering
  \includegraphics{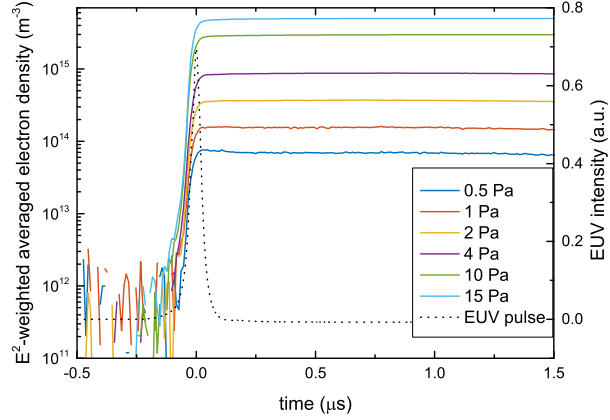}
  \caption{The electron density as a function of time for various gas pressures. The green dashed curve shows the normalized EUV intensity. The energy of the EUV pulse is \SI{53+-3}{\micro\joule}.}
  \label{fig:elec_pressure_time_zoom}
\end{figure}

\begin{figure}[!h]
  \centering
  \includegraphics{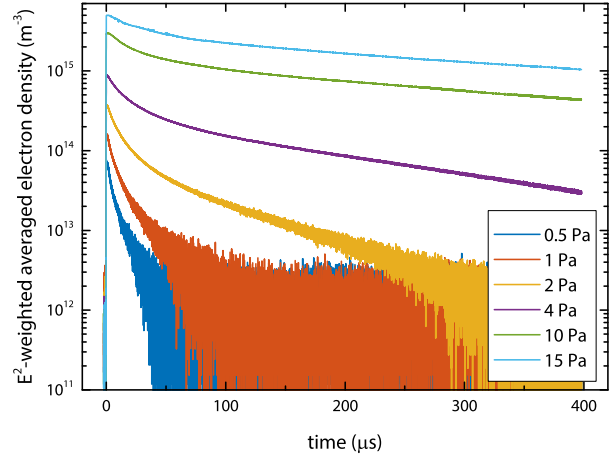}
  \caption{The electron density as a function of time for various gas pressures at a larger time scale. The energy of the EUV pulse is \SI{53+-3}{\micro\joule}.}
  \label{fig:elec_pressure_time}
\end{figure}

We measured the electron density as a function of time for various gas pressures in the range \SIrange{0.5}{15}{\pascal}. Results are shown for respectively short and long time scales in Figures~\ref{fig:elec_pressure_time_zoom}~and~\ref{fig:elec_pressure_time}. These results show that the maximum weighted and averaged electron density increases with increasing pressure. At low pressures this maximum is reached at the end of the EUV pulse, of which the temporally resolved intensity is also plotted in Figure~\ref{fig:elec_pressure_time_zoom}), while for higher pressures it is reached later. Furthermore, the results indicate that the decay time becomes longer if the pressure increases. The noise level becomes significant below a density of \SI{1e13}{\per\cubic\meter} because we approach the detection limit of our experimental set-up (\SI{\sim e12}{\per\cubic\meter}).

\begin{figure}[!h]
  \centering
  \includegraphics{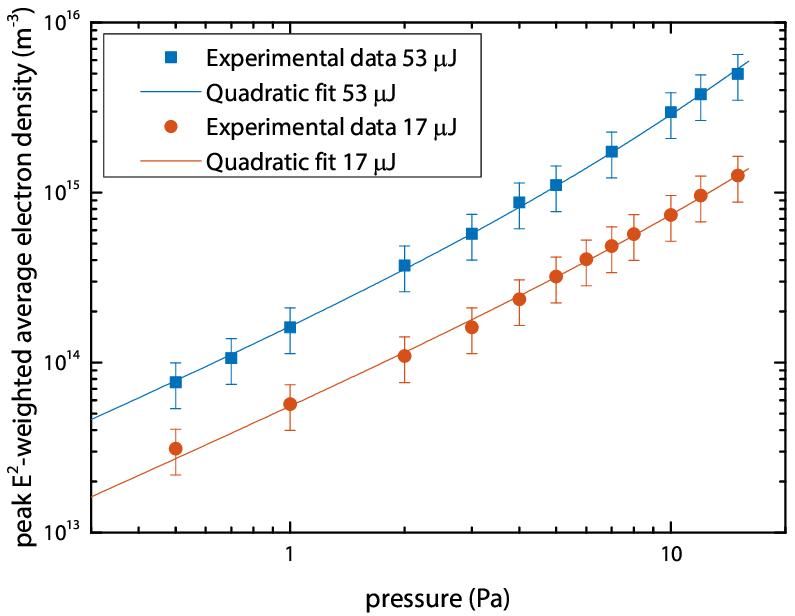}
  \caption{Maximum electron density as a function of gas pressure for two different EUV energies, i.e. \SI{53+-3}{\micro\joule} and \SI{17+-1}{\micro\joule}. The expressions of the quadratic fits are, respectively, $\bar{n}^{max}_e = \num{1.4+-0.5E13} p^2 + \num{1.5+-0.2e14} p$ and $\bar{n}^{max}_e = \num{2(1)e12} p^2 + \num{5.3+-0.9e13} p$.}
  \label{fig:max_elec_pressure}
\end{figure}

The maximum electron density reached after the EUV pulse $\bar{n}_e^{max}$ is plotted as a function of the gas pressure in Figure~\ref{fig:max_elec_pressure} for two different EUV pulse energies (\SIlist[]{17+-1;53+-3}{\micro\joule}). We could nicely fit the data with a quadratic function, as will be explained in section~\ref{sec:Discussion}. For an EUV pulse energy of \SI{53+-3}{\micro\joule} the fit has the following expression: $\bar{n}^{max}_e = \num{1.4+-0.5E13} p^2 + \num{1.5+-0.2e14} p$. The expression of the fit for an EUV pulse energy of \SI{17+-1}{\micro\joule} is: $\bar{n}^{max}_e = \num{2(1)e12} p^2 + \num{5.3+-0.9e13} p$.
	
\subsection{EUV pulse energy}
\begin{figure}[!h]
	\centering
	\includegraphics{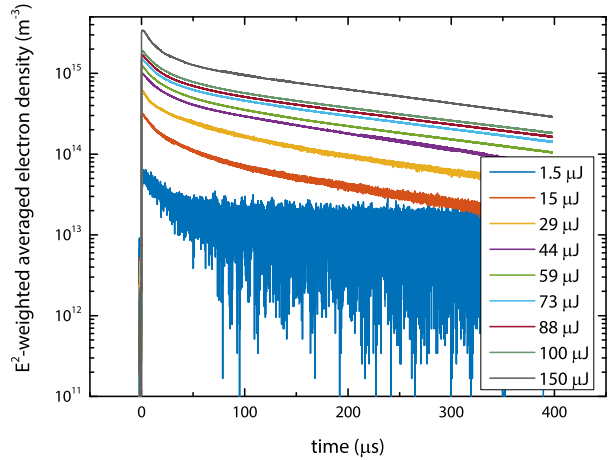}
	\caption{The electron density as a function of time for various EUV pulse energies. The pressure was \SI{5}{Pa}}
	\label{fig:elec_intensity_time}
\end{figure}

We also measured the electron density as a function of time for various EUV pulse energies. Results are shown in Figure~\ref{fig:elec_intensity_time} and demonstrate that the maximum electron density to be reached increases with increasing EUV pulse energy. Furthermore, they show that decay times do not significantly depend on the pulse energy.

\begin{figure}[!h]
	\centering
	\includegraphics{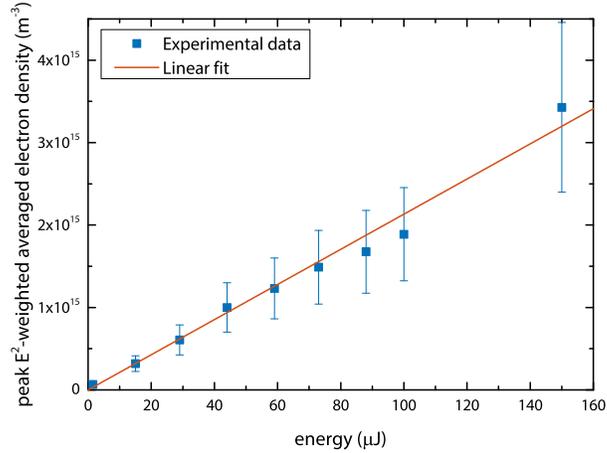}
	\caption{Maximum electron density as a function of EUV pulse energy. The fit has a slope of \SI{2.1+-0.2e13}{\per\cubic\meter\per\micro\joule} and the intercept was fixed at zero.}
	\label{fig:max_elec_intensity}
\end{figure}

The maximum electron density reached is plotted as a function of the EUV pulse energy in Figure~\ref{fig:max_elec_intensity}. We could fit the data well with a linear fit with an intercept fixed at zero. The slope of this fit is \SI{2.1+-0.2e13}{\per\cubic\meter\per\micro\joule}.

\section{Discussion and interpretation} \label{sec:Discussion} 
We have measured the electron density as a function of gas pressure and EUV pulse energy. In this section we will discuss the results in therms of relevant physical processes.

\subsection{Gas pressure} \label{subsec:pressure_discussion}
Our experimental results indicate a quadratic dependence of the maximum reached electron density on the gas pressure. To gain more insight we explore the production and loss processes of electrons in this configuration. The most relevant among these are: 
\begin{enumerate}
\item Production due to photo-ionization $P_{pi}$.
\item Production due to electron impact ionization $P_{ei}$.
\item Loss of initial high energy electrons $L_{fe}$.
\item Indirect loss due to averaging $L_{av}$.
\item Loss due to recombination $L_{rec}$.
\end{enumerate}
Next, these processes will be described briefly.

\begin{table}
\centering
\caption{Characteristic time scales for various processes in EUV induced plasmas.}
\label{tab:timescales}
\begin{tabular}{ll}
\toprule
Process						&	Typical time scale					\\
\midrule
EUV pulse length			&	\SI{100}{\nano\second}				\\
Post EUV ionization	time 	&	\SIrange{1}{10}{\micro\second}		\\
Time to reach wall			&	\SIrange{10}{100}{\micro\second}	\\
Total decay time			&	\SIrange{0.01}{1}{\milli\second}	\\
Time between EUV pulses		&	\SI{2}{\milli\second}				\\
\bottomrule
\end{tabular}
\end{table}

\begin{description}
\item[Photo-ionization] The first production process is absorption of EUV photons $\gamma$ and subsequent photo-ionization $P_{pi}$~\cite{Saito1992}:
	\begin{equation}
	\ce{$\gamma$ + Ar -> e- + Ar+}.
	\end{equation}
with a cross section at \SI{92}{\electronvolt} of \SI{1.1e-22}{\square\meter}~\cite{Saito1992} and an ionization energy of \SI{15.8}{\electronvolt}~\cite{Kramida2013} or:
	\begin{equation}
	\ce{$\gamma$ + Ar -> 2 e- + Ar^2+}.
	\end{equation}
with a cross section at \SI{92}{\electronvolt} of \SI{2.5e-23}{\square\meter}~\cite{Saito1992} and an ionization energy of \SI{43.4}{\electronvolt}~\cite{Kramida2013}. Triple ionization  is negligible under our circumstances~\cite{Saito1992}.  Due to momentum conservation, almost all excess energy is transferred to the electrons while the ions remain roughly at room temperature. So every absorbed photon creates on average 1.2 electrons with an averaged energy of \SI{66}{\electronvolt} (\SI{80}{\percent} single ionization and \SI{20}{\percent} double ionization). 

\item[Electron impact ionization] Secondly we have production due to electron impact ionization $P_{ei}$~\cite{Phelps1999,Jha2006}:
	\begin{align}
	\cee{e- + Ar &-> 2 e- + Ar+\\
				 &-> 3 e- + Ar^2+},
	\end{align}
with cross sections at \SI{66}{\electronvolt} of \SI{2.7e-20}{\square\meter}~\cite{Phelps1999} and \SI{1.3e-21}{\square\meter}~\cite{Jha2006}, respectively.

\item[Loss of high energy electrons] As mentioned before, the initial electrons have an average energy of \SI{66}{\electronvolt}. These high-energy electrons escape from the centre of the cavity, leaving the positive ions behind. As a result a potential difference is induced between the plasma in the centre and the wall. The governing electric field confines the remaining electrons in the plasma. To estimate the charge density needed to confine the plasma, we approximate the cavity with the initial plasma just after the EUV irradiation by the configuration of a coaxial cable.The core of the coax represents the positive ion column while the shield of the coax represents the cavity wall. The electric field $\vec{E}$ in a coaxial cable can be determined with Gauss's law:
	\begin{equation}
	\oint_A \vec{E}\cdot d\vec{a} = \frac{Q_{encl}}{\epsilon_0},
	\end{equation} 
where $Q_{encl}$ is the charge enclosed by surface $A$. Solving this equation for $E$ and integrating over $r$ yields the potential between the core of the core and the shield:
	\begin{equation}
	\Delta V = -\frac{\lambda_q}{2\pi r \varepsilon_0} \ln\left(\frac{R_s}{R_c}\right),
	\end{equation}
with $\lambda_q$ the charge density per unit length, $R_s$ the radius of the sheath and $R_c$ the radius of the core. If we rewrite this equation for a plasma ($\lambda_q = Zen_+ r_p^2$, $R_c=r_p$ and $R_s=R$), the potential between the ions (core) and wall (shield) reads:
	\begin{equation}
	\Delta V = -\frac{Z e n_+ r_p^2}{2\varepsilon_0} \ln\left(\frac{R}{r_p}\right),
	\end{equation}
where $Z$ is the ion charge, $n_+$ is the ion density in the core, $r_p$ is the radius of the plasma and $R$ is the radius of the cavity. We assume that the radius of the core is equal to the radius of the EUV beam (\SI{2}{\milli\meter}). The average ion charge is \num{1.2} (see previous point). From our calculation we find a necessary ion density of \SI{6E14}{\per\cubic\meter} in order to confine all electrons. To compare this to the measured electron density, we need to convert the ion density to a squared-electric-field weighted averaged ion density.  Since we know the mode of the cavity (TM$_{010}$) and the local (relative) electric fiels strength, we can convert the ion density to an averaged density using equation~\ref{eq:avDensity}, which results in an averaged ion density of \SI{6E12}{\per\cubic\meter}. Since the average ion charge is \num{1.2}, this corresponds to an averaged electron density of \SI{8E12}{\per\cubic\meter} (initially quasi-neutral plasma), as to be measured with our MCRS method. So the loss term due to the fast electrons is small compared to the measured electron density and can be safely neglected.

\item[Indirect loss due to averaging] Another loss process is related to the square-electric-field weighted averaging $L_{av}$. If the plasma expands toward the wall, the apparent averaged density decreases since the E-field decreases, i.e. electrons are sampled to a lower extent. However, the E-field in our experimental cavity is approximately constant for \SI{10}{\milli\meter} around the cavity axis (see Figure~\ref{fig:Efield}). We estimate the time it takes the plasma to expand to \SI{10}{\milli\meter} from the measurement at \SI{0.5}{\pascal}, since electron impact ionization is negligible at this pressure. The cavity-averaged density is constant for \SI{1}{\micro\second} (see Figure~\ref{fig:elec_pressure_time_zoom}), so the loss due to averaging is not significant before \SI{1}{\micro\second}. 

\item[Loss due to recombination] Since it takes even longer for the plasma to reach the wall (radius of \SI{33}{\milli\meter}) than the previously mentioned \SI{1}{\micro\second}, losses due to wall recombination $L_{rec}$ are also neglected at time scales shorter than \SI{1}{\micro\second}.
\end{description}
Typical time scales related to these processes are summarized in table~\ref{tab:timescales}.

The maximum electron density is always reached before \SI{1}{\micro\second}. Consequently, the rate equation for the electron density only includes $P_{pi}$ and $P_{ei}$:
	\begin{eqnarray}
	\frac{dn_e}{dt} &=&	P_{pi} + P_{ei} 	\underbrace{-L_{fe}-L_{av}-L_{rec}}_{=0} \nonumber	\\
					&=& \frac{1.2\sigma_{pi} n_a E_{EUV}(t)}{A_{EUV} E_{ph} \tau} +
						k_{ei} n_e n_a  \label{eq:diff_equation} \\
					&=& a(t) + b n_e,\nonumber
	\end{eqnarray}
where $\sigma_{pi}$ is the photo-ionization cross section, $n_a$ is the background density, $A_{EUV}$ is the cross section of the EUV beam, $E_{ph}$ is the photon energy, $\tau$ is the duration of the EUV pulse, $k_{ei}$ the electron impact ionization rate and
	\begin{eqnarray}
	E_{EUV}(t) =
		\begin{dcases}
		E_{EUV}, & 0<t<\tau \\
		0,       & t>\tau
		\end{dcases}
	\end{eqnarray}
where $E_{EUV}$ is the EUV pulse energy. The solution of equation~\ref{eq:diff_equation} reads (see Appendix~\ref{app:solution}):
	\begin{equation}
	n_e^{max} = \frac{1.2\sigma_{pi} n_a E_{EUV}}{A_{EUV} E_{ph}}
					\left[1 + k_{ei} n_a \left(t_m - \frac{1}{2}\tau\right)\right].
	\label{eq:max_density}
	\end{equation}
This means that the maximum reached electron density increases linearly with EUV pulse energy and quadratically with the background density. This is exactly what we have observed in the experimental results in the framework of the pressure dependence study discussed in this section; we could fit a quadratic function perfectly through our experimental data (Figure~\ref{fig:max_elec_pressure}).

The linear coefficient in the quadric fits in Figure~\ref{fig:max_elec_pressure} has a value of \SI{1.5+-0.2e14}{\per\cubic\meter\per\pascal} and \SI{5.3+-0.9e13}{\per\cubic\meter\per\pascal} for an EUV pulse energy of \SI{53}{\micro\joule} and \SI{17}{\micro\joule},  respectively. Note that, these values are valid for the squared-electric-field weighted averaged electron density. To convert these to  coefficients which are valid for the peak electron density in the centre of the cavity, we need to assume a spatial distribution for the electron density just after the EUV pulse. We assumed that during the EUV pulse this distribution is a square with a width  equal to the FWHM (full width at half maximum) of the EUV beam, i.e. \SI{4}{\milli\meter}. Since we know the mode of the cavity (TM$_{010}$), we can convert the averaged electron density to a density in the centre of the cavity using equation~\ref{eq:avDensity}. This results in linear coefficients of \SI{1.4E16}{\per\cubic\meter\per\pascal} and \SI{5e15}{\per\cubic\meter\per\pascal} respectively. These coefficients correspond to the following term (linear coefficient) in equation~\ref{eq:max_density}:
	\begin{equation}
	\alpha_t = \frac{1.2\sigma_{pi} n_a E_{EUV}}{A_{EUV} E_{ph}} 	\frac{1}{p},
	\end{equation}
which we can rewrite using the ideal gas law:
	\begin{equation}
	\alpha_t = \frac{1.2\sigma_{pi} E_{EUV}}{A_{EUV} E_{ph} k_b T_g},
	\end{equation}
where $p$ is the gas pressure, $k_b$ is the Boltzmann constant and $T_g$ the gas temperature. We can estimate this linear coefficient from the absorbed EUV pulse energy. The cross section, photon energy and EUV energy depend on the wavelength of the photon $\lambda$, so we write the previous equation as:
	\begin{equation}
	\alpha_t = \int_0^\infty \frac{1.2\sigma_{pi}(\lambda) E_{EUV}(\lambda)}{k_b T_g A_{EUV} hc/\lambda} d\lambda,
	\end{equation}
where $h$ is the Planck constant and $c$ is the speed of light. To calculate this coefficient, we need the photo-ionization cross section and the EUV energy. The cross section is taken from Saito and Suzuki~\cite{Saito1992}. The spectrally resolved EUV pulse energy is calculated from the relative EUV spectrum recorded by Kieft~\cite{Kieft2008} and a wavelength integrated energy measurement of the EUV source (\SI{53}{\micro\joule} per pulse). This results in a linear coefficient of \SI{2E16}{\per\cubic\meter\per\pascal}, which corresponds well with the experimental value (\SI{1.4E16}{\per\cubic\meter\per\pascal}). When we decrease the EUV energy by a factor 3 to \SI{17}{\micro\joule}, the theoretical coefficient also decreases by a factor 3 due to the linear dependence on the EUV pulse energy. This is also observed experimentally (a factor \num{2.8+-0.8}).

The difference in the quadratic part of the fits in Figure~\ref{fig:max_elec_pressure} when decreases the energy with a factor 3, should also be a factor of 3 since it scales with the EUV pulse energy (see equation~\ref{eq:max_density}). Experimentally we found a difference of a factor $7\pm4$, which corresponds to 3 within the margin of error.

Our experimental results (Figure~\ref{fig:elec_pressure_time_zoom}) show that at low pressures the maximum electron density is reached at the end of the EUV pulse, while at higher pressures this maximum is reached after well the EUV pulse. This difference in behaviour can be explained to the fact that at low pressures the total ionization is mainly governed by photo-ionization, hence the production of electrons stops at the end of the EUV pulse and the maximum is reached. At higher pressures electron impact ionization continues to create electrons after the EUV pulse at a significant rate.

\subsubsection{Plasma decay}
As mentioned before, a potential difference is induced between the plasma and the cavity walls. On sufficiently long time scales, this causes an ambipolar flow~\cite{Franklin2003}. Due to this flow, the plasma expands to the walls of the cavity. Since the weighting factor of the MCRS diagnostic is smaller close to the walls of the cavity, the weighted average density decreases upon plasma expansion. However, it cannot be concluded that the total density also decreases.  In general, the speed of this expansion is the ion acoustic speed ($v = \sqrt{k_B T_e/m_i}$)~\cite{Samir1983}, which only holds after the first very fast electrons escaped to the wall. Since the electron temperature decreases with time and the rate of this decrease depends on the gas pressure, we can only indicate a range of expansion speeds: $v=\SIrange[range-phrase=-]{5e2}{5e3}{\meter\per\second}$ for $\hat{T}_e= \SIrange{0.1}{10}{\electronvolt}$. So it takes \SIrange{7}{70}{\micro\second} for the plasma to reach the wall of the cavity. When the plasma reaches the wall, the electrons and ions recombine. Typically, the decay time $\tau$ of the plasma is then governed by the ambipolar flow~\cite{Luikov1968}:
	\begin{equation}
	\tau = \left[ \left(\frac{2.405}{R}\right)^2 +
				\left(\frac{\pi}{H}\right)^2 \right] ^{-1} \frac{1}{D_a},
	\label{eq:tau}
	\end{equation}
with $R$ and $H$ the radius and height of the cavity and $D_a$ the ambipolar diffusion coefficient. The equation above is only valid if an ambipolar equilibrium has been established. The ambipolar diffusion coefficient depends on the ion mobility $\mu_i$, the electron temperature $\hat{T}_e$ and ion temperature $\hat{T}_i$ (in \si{\electronvolt}):
	\begin{equation}
	D_a = \mu_i \left(\hat{T}_e + \hat{T}_i\right),
	\label{eq:Damb}
	\end{equation}
The ion mobility decreases if the pressure increases~\cite{Helm1977,Ellis1976}. This perfectly explains our experimental observation of the plasma decaying faster at lower pressures.

\subsection{EUV pulse energy}

We also studied the dependence of the electron density on the EUV pulse energy at \SI{5}{\pascal}. The experimental results (Figure~\ref{fig:max_elec_intensity}) show that the maximum electron density depends linearly on the EUV pulse energy. The slope of squared-electric-field weighted averaged electron density against the pulse energy is \SI{2.1+-0.2e13}{\per\cubic\meter\per\micro\joule}. This linear dependence is also predicted by theory (equation~\ref{eq:max_density}).

The experiments (Figure~\ref{fig:elec_intensity_time}) show that the decay time does not depend on the EUV pulse energy. This is also clear from equations~\ref{eq:tau}~and~\ref{eq:Damb}.

\section{Conclusions}

We experimentally explored the electron density evolution in an EUV induced plasma in argon using microwave cavity resonance spectroscopy. The dependence of the electron density on the EUV pulse energy and gas pressure was investigated. We found that the maximum reached electron density depends linearly on the EUV pulse energy as also expected from theory. Also, our experimental results showed that the maximum reached electron density depends quadratically on the gas pressure. The linear term can be attributed to photo-ionization while the quadratic term is due to electron impact ionization. Finally, we found that the decay time of the EUV-induced plasma, where plasma recombination on the cavity walls is dominant, becomes longer if the pressure increases, which is explained by an increase of the ion mobility, hence the ambipolar diffusion coefficient, with increasing pressure.

To gain more insight in the decay of the plasma, the spatial distribution of the plasma density as function of time should be determined. This can for instance be done by imaging the plasma. Furthermore, it is recommended to study the plasma in  gasses with a smaller absorption cross section, such as hydrogen or helium, as used in lithography tools.

\section{Appendix}\label{app:solution}
The general solution of the equation:
	\begin{eqnarray}
	\frac{dn_e}{dt} &=& \frac{1.2\sigma_{pi} n_a E_{EUV}(t)}{A_{EUV} 	
						E_{ph} \tau} + k_{ei} n_e n_a \\
					&=& a(t) + b n_e.
	\end{eqnarray}
is~\cite{Adams2010}:
	\begin{equation}
	n_e = e^{-M(t)} \int e^{M(t)} a(t) dt,
	\end{equation}
where $M(t)$ is the integrating factor:
	\begin{equation}
	M(t) = \int b dt.
	\end{equation}
With the initial condition $n_e(0)=0$, this leads to the following solution:
	\begin{equation}
	n_e(t) =
		\begin{dcases}
		\frac{a^*}{b}\left(e^{bt} - 1\right), & 0<t<\tau \\
		\frac{a^*}{b}\left(e^{bt} - e^{b(t-\tau)}\right), & t>\tau
		\end{dcases}
		\label{eq:density_evo}
	\end{equation}
where $a^*=a(0<t<\tau)$.
The maximum density in Figure~\ref{fig:elec_pressure_time_zoom} is reached at $t_m\geq\tau$, so the equation for the maximum density is:
	\begin{equation}
	n_e^{max} =
		\frac{a^*}{b}\left(e^{bt_m} - e^{b(t_m-\tau)}\right).
		\label{eq:density_max}
	\end{equation}
The term $bt_m$ in the exponent is smaller than 1, so we can simplify the equation with a Taylor expansion:
	\begin{eqnarray}
	n_e^{max} &=& \frac{a^*}{b}\left(1+bt_m+\frac{1}{2}b^2t_m^2 - 1 - b(t_m-\tau) -
					\frac{1}{2}b^2(t_m-\tau)^2\right)\\
			  &=& a^*\tau\left(1+b(t_m-\frac{1}{2}\tau)\right).
	\end{eqnarray}
If we substitute $a^*$ and $b$ we get the following expression for the maximum density:
	\begin{equation}
	n_e^{max} = \frac{1.2\sigma_{pi} n_a E_{EUV}}{A_{EUV} E_{ph}}
					\left[1 + k_{ei} n_a \left(t_m - \frac{1}{2}\tau\right)\right].
	\end{equation}

\section*{Acknowledgements}
The authors would like to acknowledge ASML for their financial support and the opportunity to use its EUV sources.	
	
\section*{References}
\bibliographystyle{iopart-num}
\bibliography{MyPapers-ArgonDensityPressure}

\end{document}